\documentclass[aps,prl,notitlepage,twocolumn,superscriptaddress,nolongbibliography]{revtex4-1}
\usepackage{amsmath}
\usepackage{amssymb}
\usepackage{graphicx}                           
\usepackage{dcolumn}                           
\usepackage{bm}                                     
\usepackage[colorlinks=true,linkcolor=blue,anchorcolor=blue, citecolor=blue,urlcolor=blue]{hyperref}

\begin{document}

\preprint{APS/123-QED}

\title{General Electronic Structure Calculation Method for Twisted Systems}
\author{Junxi Yu}
\thanks{These authors contributed equally to this work.}
\affiliation{Centre for Quantum Physics, Key Laboratory of Advanced Optoelectronic Quantum Architecture and Measurement (MOE), and Beijing Key Lab of Nanophotonics and Ultrafine Optoelectronic Systems, School of Physics, Beijing Institute of Technology, Beijing 100081, China}
\author{Shifeng Qian}
\thanks{These authors contributed equally to this work.}
\affiliation{Centre for Quantum Physics, Key Laboratory of Advanced Optoelectronic Quantum Architecture and Measurement (MOE), and Beijing Key Lab of Nanophotonics and Ultrafine Optoelectronic Systems, School of Physics, Beijing Institute of Technology, Beijing 100081, China}
\affiliation{Anhui Province Key Laboratory for Control and Applications of Optoelectronic Information Materials, Department of Physics, Anhui Normal University, Anhui, Wuhu 241000, China}
\author{Cheng-Cheng Liu}
\email{ccliu@bit.edu.cn}
\affiliation{Centre for Quantum Physics, Key Laboratory of Advanced Optoelectronic Quantum Architecture and Measurement (MOE), and Beijing Key Lab of Nanophotonics and Ultrafine Optoelectronic Systems, School of Physics, Beijing Institute of Technology, Beijing 100081, China}

\begin{abstract}
In recent years, two-dimensional twisted systems have gained increasing attention. However, the calculation of electronic structures in twisted material has remained a challenge. 
To address this, we have developed a general computational methodology that can generate twisted geometries starting from monolayer structure and obtain the precisely relaxed twisted structure through a machine learning-based method. Then the electronic structure properties of the twisted material are calculated using tight-Binding (TB) and continuum model methods, thus the entire process requires minimal computational resources. In this paper, we first introduce the theoretical methods for generating twisted structures and computing their electronic properties. We then provide calculations and brief analyses of the electronic structure properties for several typical two-dimensional materials with different characteristics. This work serves as a solid foundation for researchers interested in studying twisted systems.
\end{abstract}

\maketitle

\section{$\rm{\uppercase\expandafter{\romannumeral1}}$. introduction}

With the pioneering work of Geim et al.\cite{novoselov_electric_2004}, 
who successfully isolated single-atom layer graphene from graphite, 
two-dimensional(2D) material came into the spotlight. 
Subsequently, other 2D material, 
such as boron nitride (BN) \cite{dean_boron_2010}, transition metal dichalcogenides (TMDs) \cite{mak_atomically_2010}, 
and black phosphorene \cite{li_black_2014,samuel_reich_phosphorene_2014}, were also successfully isolated. 
Two-dimensional materials exhibit exceptional properties such as high carrier mobility and unique energy spectra, 
which have gradually captured the attention of researchers \cite{butler_progress_2013, fiori_electronics_2014,wang_one_dimensional_2013,xia_two-dimensional_2014}. 
Due to their relatively simple crystal structures, a subset of 2D material, notably graphene, 
has become a testing ground for various theoretical studies.

Stacking 2D material together in various configurations leads to the emergence of distinct properties.
In particular, in bilayer 2D material, 
introducing a relative rotation angle ($\theta$) between the upper and lower layers around a specific axis breaks the original symmetry,
giving rise to a long-period moir\'e lattice \cite{lopes_dos_santos_continuum_2012}. 
A moir\'e primitive cell often encompasses multiple original primitive cells, and its size depends on the rotation angle; typically, 
smaller rotation angles correspond to larger moir\'e unit cell sizes. 
This rotation-induced change significantly impacts the energy band structure of the system\cite{tarnopolsky_origin_2019,kariyado_flat_2019,bao_deep-learning_2024,cea_twists_2019,ma_topological_2021,carr_exact_2019,qian_stable_2023,yu_origin_2023}.
Initially, twisted bilayer graphene (TBG) attracted broad interest for its topologically non-trivial flat band with extremely narrow bandwidth \cite{bistritzer_moire_2011,po_faithful_2019,ahn_failure_2019}. 
Additionally, TBG exhibits strong electron-electron interactions, making it an excellent platform for investigating electronic correlation effects \cite{xie_spectroscopic_2019}. 
Furthermore, TBG boasts a rich phase diagram, featuring correlated insulators, Chern insulators, and unconventional superconductors, among others \cite{cao_correlated_2018,serlin_intrinsic_2020,cao_unconventional_2018,yankowitz_tuning_2019,lu_superconductors_2019,cao_ab_2021,liu_chiral_2018}, 
and possesses unique optical and acoustic properties \cite{moon_optical_2013,gao_tunable_2020,liu_anomalous_2020,cocemasov_phonons_2013,wu_theory_2018}. 
Consequently, it has emerged as a prominent material within the graphene family.
Another class of twisted material that has garnered widespread attention is the twisted transition metal dichalcogenide (TMD) system. 
In this system, researchers have observed continuous Mott phase transitions and quantum criticality phenomena \cite{li_continuous_2021}. 
More recently, the observation of the fractional quantum Hall effect and superconductivity in the twisted bilayer TMD system has ignited a new research wave \cite{cai_signatures_2023,guo2024superconductivity}. 
Additionally, recent studies have found that twist can induce altermagnetism with impressive physical properties in 2D materials \cite{liu2024twisted}.

\begin{figure*}
	\includegraphics[width=\textwidth]{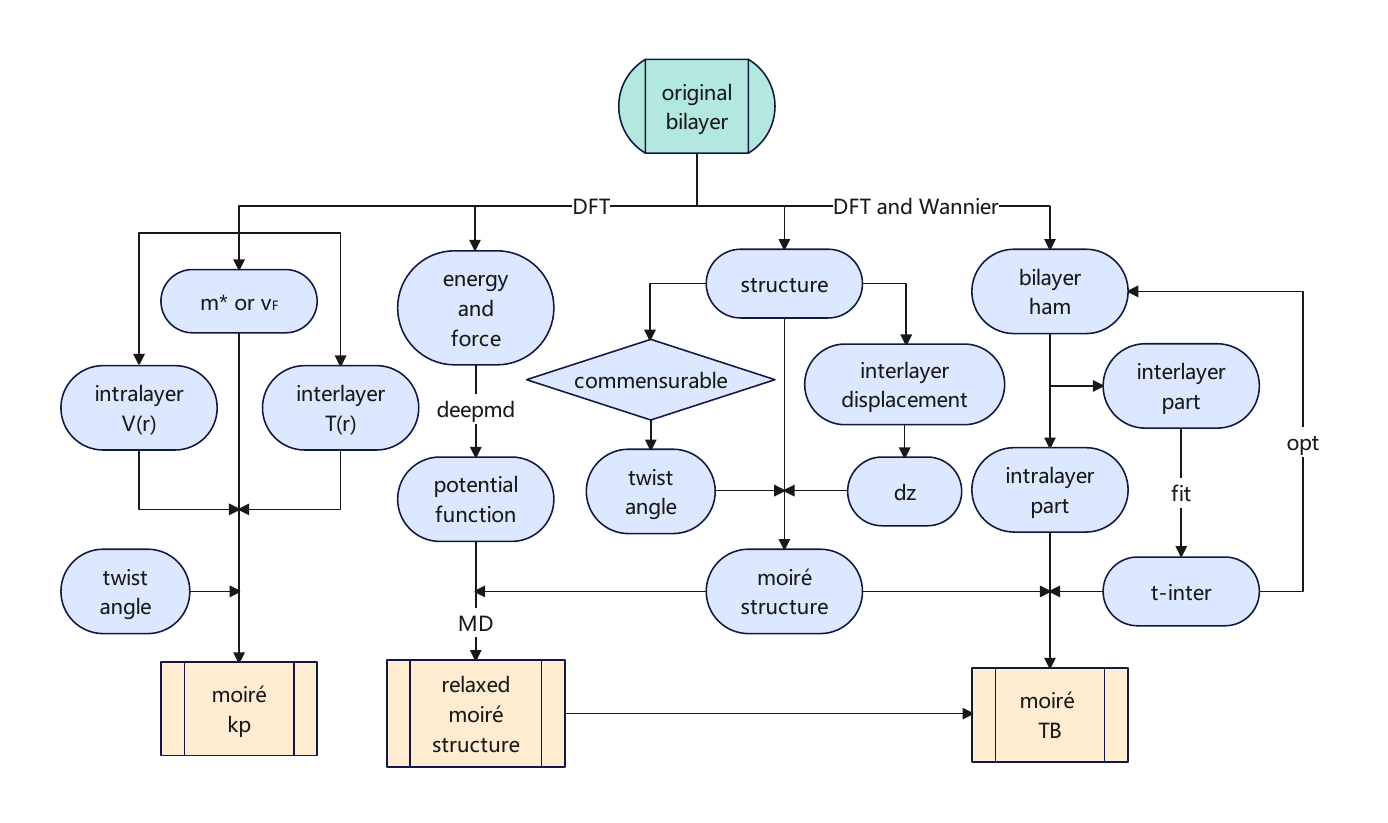}
	\caption{The overall workflow of the general method for calculating the 
	electronic structures of twisted materials start from density functional theory (DFT) calculations of the original bilayer material. 
	It eventually leads to obtaining the relaxed moir\'e structure, moir\'e TB Hamiltonian, 
	and moir\'e k·p Hamiltonian. 
	This comprehensive approach allows us to investigate the properties of twisted systems from multiple perspectives.}
\end{figure*}

Overall, twist, as a new degree of freedom in the study of material properties, with the increasing maturity of material preparation methods and 
micro-nano processing technology has become increasingly important. 
However, despite significant advancements in research methods for twisted systems, 
the approaches for calculating their geometric structures and electronic properties remain unstructured and have numerous issues.
Therefore, in addressing such issues,
we have developed a general method that allows for the investigation of 
the geometric structure and electronic properties of twisted systems from multiple perspectives.
Using this method, we can derive the moir\'e lattice structure and electronic properties of twisted systems of any 2D material.
We can then choose to investigate their electronic properties using either tight-binding (TB) or $\bm{k}\cdot\bm{p}$ continuum model methods. 
The entire process requires only minimal computational resources.
This provides a solid platform for those interested in studying the electronic structures of twisted systems.

This paper is structured as follows. In the Methodology section, 
we first provide a concise analysis of the method employed for generating the geometric structures of twisted systems, 
and we present the specific approaches used for investigating the electronic properties of twisted systems through both the TB model and the effective continuum model.
In the results section, we apply our general method to several interesting 2D systems with distinct characteristics that have garnered significant interest within the research community.
We conduct brief analyses and studies on their geometric structures and electronic properties, shedding light on their unique features.
In the Discussion section, we offer insights into the functionalities of our method and our future directions for enhancing and expanding the method.

\section{$\rm{\uppercase\expandafter{\romannumeral2}}$. methodology}
\subsection{A. Twist geometry}
A single-layer 2D material is stacked on top of another 2D material, 
and the two layers have a relative rotation angle $\theta$. 
At certain specific angles, a special periodicity occurs, known as a moir\'e pattern.
This is what we commonly refer to as twisted 2D materials.

For simplicity, we discuss homogeneous twisted material here, 
assuming perfect alignment between the upper and lower layers. 
We rotate them around an axis perpendicular to the plane as shown in Fig. 2, 
considering upper-layer and lower-layer rotations of $-\frac{\theta}{2}$ and $\frac{\theta}{2}$, respectively. 
But only when the twist angle is certain discrete specific values, 
the twisted structure has complete lattice periodicity, which we call "commensurate". 
Below we will introduce how to find commensurate twist angles.
We denote the lattice vectors of the upper and lower layers as $\bm{a}_{i}^{t}$ and $\bm{a}_{i}^{b}$, respectively. 
If there are integers $m_1, m_2, n_1, n_2$ not all zero such that
\begin{equation}
	R(-\theta/2)(m_1 \bm {a}_{1}^{t} + m_2 \bm {a}_2^{t}) = R(\theta/2)(n_1 \bm {a}_{1}^{b} + n_2 \bm {a}_{2}^{b}),
\end{equation}
then the system is commensurate at the twist angle $\theta$.
In other words, the system exhibits complete long-range periodicity only 
when the integer multiples of the lattice vectors of the upper and lower layers differ by a rotation operation.
This equation can also be expressed in matrix form
\begin{equation}
	L_t
\left(
	\begin{array}{c}
		m_1 \\
		m_2
	\end{array}
\right)
=
R(\theta)L_b
\left(
	\begin{array}{c}
		n_1 \\
		n_2
	\end{array}
\right),
\end{equation}
here, $L_t$ and $L_t$ represent the lattice vector matrices of the upper and lower layers, respectively. It can be derived as follows:
\begin{equation}
\left(
\begin{array}{c}
	m_1 \\
	m_2
\end{array}
\right)
=
(L_t)^{-1}R(\theta)L_b
\left(
	\begin{array}{c}
		n_1 \\
		n_2
	\end{array}
\right).
\end{equation}
Because $m_1, m_2, n_1, n_2$ are integers, $(L_t)^{-1}R(\theta)L_b$ must be a rational matrix \cite{lopes_dos_santos_continuum_2012}.
Based on this requirement, we can derive a series of Diophantine equations. By solving these equations, 
we can determine whether the system is commensurate at a given rotation angle $\theta$.
By utilizing Eq. (3), we can identify and select suitable commensurate twist angles. 

At commensurate twist angles, we can determine the lattice structure of the moir\'e system using the following method.
The relative displacement of points originally aligned in the upper and lower layers after rotation can be expressed as:
\begin{equation}
    \bm{\tau}(\bm{r}) = R(\frac{\theta}{2})\bm{r} - R(-\frac{\theta}{2})\bm{r},
\end{equation}
here, $\bm{r}$ is the position vector of the atom at this site measured from the origin before rotation.
The above formula can simplified to the following form
\begin{equation}
    \bm{\tau}(\bm{r}) = 2sin\frac{\theta}{2}\bm{e}_{z}\times \bm{r},
\end{equation}
where $\bm{e}_{z}$ is the unit vector along the axis of twist.
When the vector difference between two originally overlapping atoms due to the rotation is equal to the integer linear combination of the original bilayer lattice vector,
that is, when $\bm{\tau}(\bm{r})=l_{1}\bm{a_{1}}+l_2\bm{a_2}$, the system
has complete lattice periodicity, where $l_1, l_2$ are positive integers. 
Through the above formula, we explore how to obtain the moir\'e lattice vector of the twisted structure.
When four integers are not all zero such that $l_1, l_2, N_1, N_2$ make the four vectors $\bm{a}_1, \bm{a}_2, \bm{A}_1, \bm{A}_2$ satisfy the following relationship, 
the system will have complete long-range periodicity, 
where $\bm{a_i}$ and $\bm{A_i}$ are lattice vector for original bilayer and its twisted structure, respectively.
\begin{figure}[t]
	\centering
	\includegraphics[scale=0.3]{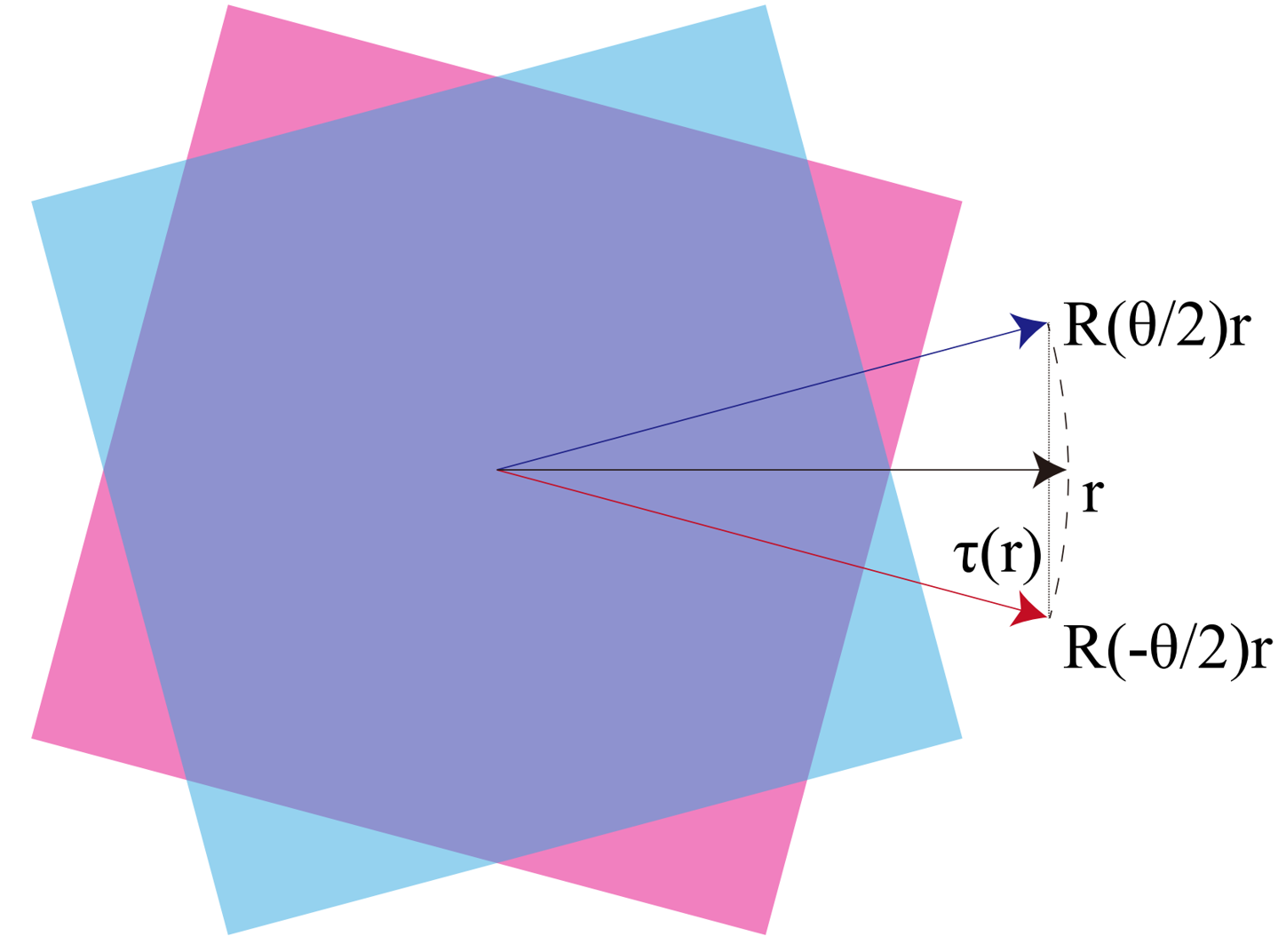}
	\caption{Two layers of two-dimensional material are stacked and rotated $\theta$, resulting in the change of any position vector in the plane.}
\end{figure}
\begin{equation}
    l_{1}\bm{a_{1}}+l_{2}\bm{a_{2}} = 2sin\frac{\theta}{2}\bm{e}_{z}\times [N_1\bm{A}_1+N_2\bm{A}_2].
\end{equation}
When the angle is small, we have $l_i=1, l_j=0, N_i=1, N_j=0$, where $i=1,2$, so there can be a simple relationship
\begin{equation}
    A_i = \frac{ \bm{a}_i\times \bm{e}_z}{2\sin\frac{\theta}{2}}.
\end{equation}
Thus, we have mastered the method to obtain the lattice structure of arbitrary twisted systems. Specific examples will be presented in the Results section.

\subsection{B. Twisted Tight-Binding model}
Now, we introduce how to construct the Hamiltonian of a twisted system within the framework of the TB method.
As shown in Fig. 1, to obtain the TB Hamiltonian of a twisted system, we need to gather three pieces of information: 
the moir\'e lattice structure, the intralayer Hamiltonian, and the interlayer coupling.

Using the method introduced in the Sec.$\rm{\uppercase\expandafter{\romannumeral2}}$A, 
we can obtain the lattice structure of the twisted system and directly apply it to the TB calculations. 
However, if higher precision is required for the electronic properties of the twisted system, 
the effects of lattice relaxation must be considered. 
Moir\'e structure relaxation can be divided into in-plane relaxation and out-of-plane relaxation. 
Typically, out-of-plane relaxation has a significant impact on the electronic structure of the system. 
For instance, in twisted graphene systems, out-of-plane relaxation is crucial for the formation of isolated flat bands\cite{nam_lattice_2017}.

Therefore, we start with the assumption of a local stacking approximation and consider this out-of-plane relaxation by introducing an interlayer spacing function 
$d_z(\bm{r})$ that varies with x and y.
We can accurately determine this interlayer spacing function 
$d_z(\bm{r})$ through first-principles calculations. Initially, 
we define the in-plane interlayer displacement in the original bilayer structure
\begin{equation}
	\bm{\tau} = \tau_1 \bm{a_1} + \tau_2 \bm{a_2},
\end{equation}
where $\bm{a_1}$ and $\bm{a_2}$ is the lattice vectors of the original bilayer structure,
$\tau_1$ and $\tau_2$
are the interlayer relative displacements in terms of lattice vectors, with values ranging from 0 to 1,
and divide this range into 40 intervals. By traversing each value of 
$\tau_1$ and $\tau_2$, we obtain 1600 structures with different in-plane interlayer displacements.
For each structure, we perform van der Waals-corrected structural relaxation to obtain precise interlayer spacings.
Fitting these interlayer spacing data on the xy plane allows us to derive an accurate interlayer spacing function $d_z(\bm{r})$.
However, this method requires considerable computational resources. 

For this issue, we have a simpler solution, which is to directly obtain the form of 
$d_z(\bm{r})$ through symmetry analysis, and then use one or two relaxed structures to determine its parameters.
For common structures, such as the hexagonal lattice of 2D materials, 
$d_z(\bm{\tau})$ takes the following form\cite{koshino_maximally_2018}:
\begin{equation}
		\begin{aligned}
		d_z(\bm{\tau}) &= c_0 + 2c_1[(\cos(\bm{G}_1 \cdot \bm{\tau})) + \\
		&\cos(\bm{G}_2 \cdot \bm{\tau}) + \cos(\bm{G}_1 \cdot \bm{\tau} + \bm{G}_2 \cdot \bm{\tau})],
		\end{aligned}
\end{equation}
where $c_0 = (d_{AA} + 2 d_{AB})/3, c_1 = (d_{AA} - d_{AB})/9$ and $\bm{G}_1,\bm{G}_2$ are the reciprocal lattice vectors of the original bilayer structure, and 
$d_{AA},d_{AB}$ are the interlayer spacings of the bilayer structure when stacked in AA and AB configurations, respectively.
For square or rectangular lattices of 2D materials, 
$d_z(\bm{\tau})$ takes the following form:
\begin{equation}
	d_z(\bm{\tau}) = c_0 + 2c_1[(\cos(\bm{G}_1 \cdot \bm{\tau})) + \cos(\bm{G}_2 \cdot \bm{\tau})],
\end{equation}
with $c_0 = (d_{AA} + d_{AB})/2, c_1 = (d_{AA} - d_{AB})/8$.
For more complex structures, 
the interlayer spacing function can be obtained by making some transformations from Eq. (9) or Eq. (10) 
according to the atomic positions of the original bilayer structure.
Therefore, we only need to calculate $d_{AA},d_{AB}$ to obtain $d_z(\bm{r})$ that meets the accuracy requirements.

The reciprocal lattice vectors 
$\bm{G}_i$ of the original bilayer structure and those 
$\bm{G}_i^M$ of the twisted structure are related as follows:
$\bm{G}_i^M=2\sin\frac{\theta}{2} \bm{G}_i  \times \bm{e}_z$,
and combining it with Eq. (5) we can get
\begin{equation}
	\bm{G}_i \cdot \bm{\tau} =  \bm{G}_i^M \cdot \bm{r},
\end{equation}
which will be used repeatedly in the following sections.
By substituting Eq. (11) into Eqs. (9) and (10), we can rewrite them as:
\begin{subequations}
    \begin{align}
        d_z(\bm{r}) &= c_0 + 2c_1[(\cos(\bm{G}_1^M \cdot \bm{r})) + \cos(\bm{G}_2^M \cdot \bm{r}) + \nonumber \\
        &\cos(\bm{G}_1^M \cdot \bm{r} + \bm{G}_2^M \cdot \bm{r})], \\
        d_z(\bm{r}) &= c_0 + 2c_1[(\cos(\bm{G}_1^M \cdot \bm{r})) + \cos(\bm{G}_2^M \cdot \bm{r})].
    \end{align}
\end{subequations}
These equations represents the function describing the variation of interlayer spacing with position in the twisted structure. 
Thus, we now understand how to introduce out-of-plane relaxation in twisted structures with minimal computational effort.
Meanwhile, the in-plane relaxation is discussed in Section $\rm{\uppercase\expandafter{\romannumeral2}}$.D.

Regarding the intralayer part of the Hamiltonian, 
the common assumption is that the intralayer hopping in a bilayer structure, 
compared to a monolayer structure, is only influenced by the overall potential energy of the other layer. 
This influence is very small and can be considered independent of interlayer displacement or rotation between the two layers. 
Therefore, we assume that the intralayer part of the Hamiltonian for the original bilayer structure is equal to the intralayer Hamiltonian of the twisted system. 
We can simply use DFT calculations to compute the properties of the original bilayer structure and perform Wannier interpolation to obtain the TB Hamiltonian of the original bilayer structure. 

The final and most crucial piece of information needed for the twisted TB model is the interlayer coupling. 
We assume that the interlayer coupling in the twisted system is real and decays exponentially with distance. 
For any two orbitals located in different layers, their hopping can be written as follows:
\begin{equation}
	t(\bm{r}) = h_{0} exp(-\frac{l(r_z-d_0)^2}{r_0^2}) exp(-\frac{r_x^2 + r_y^2}{r_0^2}).
\end{equation}
Here, 
$h_0$ refers to the coupling strength when the interlayer displacement 
$\bm{\tau}$ aligns these two orbitals in the original bilayer structure, and 
$d_0$ is the distance between them at that alignment. 
$\bm{r}$ represents the distance between the two orbitals in the twisted structure, where 
$r_x,r_y,r_z$ are its components in the 
$x,y,z$ directions, respectively, and 
$l=\pm 1$ is determined by the sign of 
$r_z-d_0$.

In Eq. (13), we envision atoms as regular spherical bodies, hence 
$r_0$ serves as a decay coefficient in all three directions. For any two different-layer orbitals in the original bilayer structure, we only need to determine three parameters 
$h_0, d_0, r_0$, to establish their coupling in the twisted structure. By applying different interlayer displacement operations to the original bilayer structure to get three different structures, 
and performing DFT calculations and Wannier interpolation for these three structures, we can precisely obtain three distinct 
$t(\bm{r})$. Viewing the equation above as a system of three equations with three unknowns 
$h_0, d_0, r_0$, solving this system yields the parameters we need.
 
We have outlined methods for obtaining three pieces of information required for a twisted system within the TB framework: 
lattice structure, intralayer Hamiltonian, and interlayer coupling. 
By this, we can construct the Hamiltonian of a twisted system in the TB model.

\subsection{C. Twisted Continuum model}

For relatively small twist angles, where there are more atoms within one unit cell of the twisted system, 
our focus often lies in the electronic properties within a low-energy range near a specific $\bm{K}$ point in reciprocal space. 
In such scenarios, it is more appropriate to employ the twisted continuum model method. 
To build an accurate twisted continuum model, we also need two pieces of information: 
the intralayer part of the Hamiltonian and the interlayer coupling.
All the parameters required can be obtained by DFT calculation with minimal computational effort.

Firstly, for the target system, we acquire the overall band structure properties of the original bilayer material. 
We focus on a specific  $\bm{K}$ point in the reciprocal space of the original bilayer structure, typically a point of extremum for valleys, 
and calculate the band structure properties near this  $\bm{K}$ point through DFT calculations. 
This allows us to determine the effective mass $m^*$ or Fermi velocity $v_F$.
The two correspond to the quadratic dispersion and linear dispersion of the band respectively.
With this information, we can then determine the intralayer part of the Hamiltonian for the twisted system.
\begin{subequations}
    \begin{align}
        h(\bm{k}) &= -\frac{\hbar^2 (\bm{k}-\bm{K}_l)^2}{2m^*}, \\
        h(\bm{k}) &= -\hbar v_F(\bm{k}-\bm{K}_l) \cdot (\sigma_x, \sigma_y),
    \end{align}
\end{subequations}
where $l$ is the layer index, 
$\sigma_i$ represents the ith Pauli matrix,
Next, we still start from the original bilayer structure, and its Hamiltonian has the following form
\begin{equation}
	H(\bm{\tau})
=
\left(
	\begin{array}{cc}
		h_+(\bm{k}) + U_+(\bm{\tau}) & T(\bm{\tau}) \\
		T^{\dagger}(\bm{\tau}) & h_-(\bm{k}) + U_-(\bm{\tau})
	\end{array}
\right),
\end{equation}
where $T(\bm{\tau})$ represents the interlayer coupling.
While $U(\bm{\tau})$ denotes the impact of interlayer displacement on the intralayer Hamiltonian, 
this term is very small and can be neglected in the vast majority of cases.
The specific forms of both two terms are determined by lattice symmetry and the BZ points of interest. 
For instance, in the most common scenario, the $\bm{K}$ point in the BZ of a hexagonal lattice, 
they take the following form\cite{wu_topological_2019}:
\begin{equation}
    T(\bm{\tau}) = \delta (1 + e^{i\bm{G}_1 \cdot \bm{\tau}} + e^{i(\bm{G}_1 + \bm{G}_2) \cdot \bm{\tau}}),
\end{equation}
\begin{equation}
	\begin{aligned}
		U_l(\bm{\tau}) &= 2\xi [\cos (\bm{G}_1 \cdot \bm{\tau} + l\phi)  \\
		 &+\cos (\bm{G}_2 \cdot \bm{\tau} + l\phi)  \\
		 &+\cos ((\bm{G}_1+\bm{G}_2) \cdot \bm{\tau} + l\phi)].
	\end{aligned}
\end{equation}
Where $l$ is the layer index, $\delta$, $\xi$, and $\phi$ are parameters to be determined through DFT calculations. In the case of an orthorhombic lattice, considering the  $\Gamma$ point, the forms are as follows:
\begin{equation}
    T(\bm{\tau}) = \delta_0 + \delta_1e^{i\bm{G}_1 \cdot \bm{\tau}} + \delta_2e^{i\bm{G}_2 \cdot \bm{\tau}}, 
\end{equation}
\begin{equation}
	\begin{aligned}
    U_l(\bm{\tau}) &= 2\xi_1\cos (\bm{G}_1 \cdot \bm{\tau} + l\phi_1) \\
	&+ 2\xi_2\cos (\bm{G}_2 \cdot \bm{\tau} + l\phi_2).
	\end{aligned}
\end{equation}
Here, $\delta_i$, $\xi_i$, and $\phi_i$ are also parameters to be determined through DFT calculations. Depending on the parameters, the equation above can represent either a continuum model for a square lattice or a rectangular lattice. The parameters in the coupling term can be specifically determined 
by performing DFT calculations on the original bilayer structure with different interlayer displacements $\bm{\tau}$.

Next, we rotate the upper layer by $-\theta/2$, and the lower layer by $\theta/2$.
Using Eq. (11), we get $\bm{k} \rightarrow -i\bm{\nabla}, \bm{\tau} \rightarrow \bm{r}$\cite{jung_ab_2014}, so
\begin{equation}
	H
=
\left(
	\begin{array}{cc}
		h_{+}(-i\bm{\nabla}) + U_+(\bm{r}) & T(\bm{r}) \\
		T^{\dagger}(\bm{r}) & h_{-}(-i\bm{\nabla}) + U_-(\bm{r})
	\end{array}
\right).
\end{equation}
In this way, we obtain the Hamiltonian of the continuum model in the twisted system, and the band structure of the twisted system can be obtained by the plane wave expansion method.

\subsection{D. Lattice relaxation}
In the twisted system, relaxation is also a widely concerned issue, which has important contributions to the generation of isolated flat bands, band topological properties, twist phonons, and excitons
\cite{nam_lattice_2017,xie_lattice_2023,liu_moire_2022,tran_evidence_2019,jia_moire_2023}. 
As mentioned earlier, for different symmetries, we can simulate the out-of-plane relaxation of the system, that is, fluctuations, through a simple periodic function, 
but this method is powerless for in-plane relaxation. 
Therefore, how to obtain an accurate fully relaxed twisted structure without large-scale DFT calculations has become a new problem. 
Thus, we now introduce a method based on machine learning and molecular dynamics to obtain a completely accurate relaxed moir\'e structure with extremely low computational effort. 
As shown in Fig. 1, we still start from the original bilayer structure, obtain the structure under different interlayer displacements, 
obtain its atomic potential energy and force field information by the DFT calculation, 
and generate the interaction potential function by machine learning methods, 
and use this to perform molecular dynamics simulation on the unrelaxed target twisted structure to obtain a fully relaxed moir\'e structure.

\section{$\rm{\uppercase\expandafter{\romannumeral3}}$. results}
\subsection{A. Graphene}
Graphene is the most extensively studied 2D material and also the cornerstone of the twisted electronics family.
We first applied our method to graphene to demonstrate its effectiveness and accuracy.

First, we introduce the twisted geometry of graphene. 
Monolayer graphene has a hexagonal lattice with two inequivalent carbon atoms, 
labeled as A and B, within the unit cell, its lattice vector is $\bm{a}_1 = a_0(1, 0), \bm{a}_2 = a_0(\frac{1}{2}, \frac{\sqrt{3}}{2})$,
Therefore, the factor in Eq. (3) can be written as
\begin{equation}
	(L_b^T)^{-1}R(\theta)L_t^T 
= 
\left(
	\begin{array}{cc}
		\cos \theta - \frac{1}{\sqrt{3}} \sin \theta & -\frac{2}{\sqrt{3}}\sin \theta \\
		\frac{2}{\sqrt{3}}\sin \theta & \cos \theta + \frac{1}{\sqrt{3}} \sin \theta
	\end{array}
\right).
\end{equation}
We require that each term in the matrix be a rational number
\begin{equation}
	\frac{1}{\sqrt{3}} \sin \theta = \frac{l_1}{l_3}, \cos \theta = \frac{l_2}{l_3},
\end{equation}
here $l_1, l_2$ and $l_3$ can be all positive integers, resulting in the Diophantine equation:
\begin{equation}
	3l_1^2+l_2^2=l_3^2.
\end{equation}
The solution of this equation can be described by two positive integers $q,p$
\begin{equation}
	l_1=2qp
,l_2=3q^2-p^2
,l_3=3q^2+p^2,
\end{equation}
therefore, we know that the commensurate twist angle satisfies the following relationship
\begin{equation}
	\theta = \arccos (\frac{3q^2-p^2}{3q^2+p^2}).
\end{equation}
We can see that every commensurate twist angle can be described by a set of positive integers.

\begin{figure}[t]
	\centering
	\includegraphics[scale=0.43]{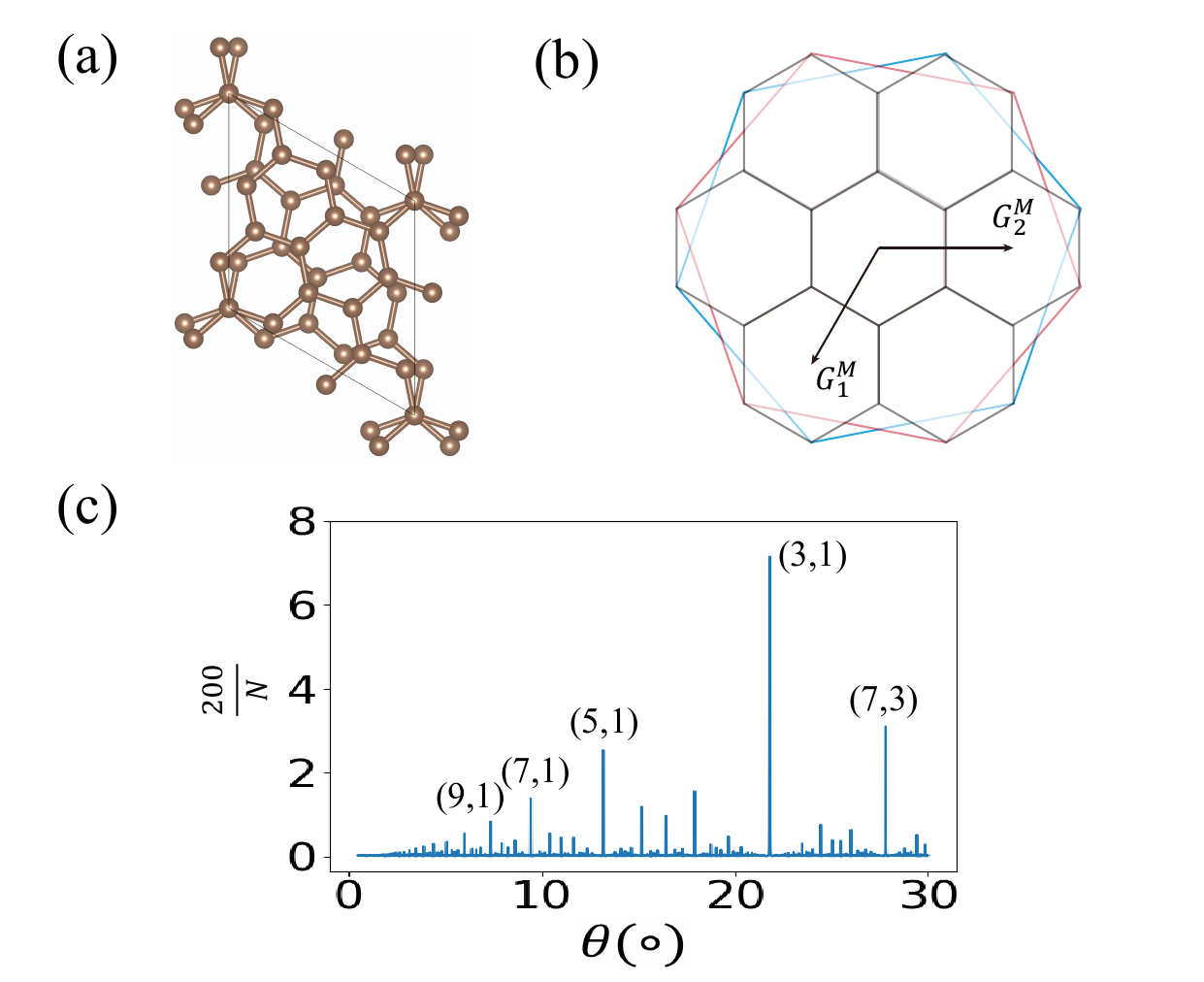}
	\caption{(a)The lattice structure of $21.79^\circ$ TBG. (b) BZ folding of $21.79^\circ$ TBG, the red(blue) hexagon represents the BZ of the original bilayer structure rotated by an angle of $-\frac{\theta}{2}$($\frac{\theta}{2}$), and the black hexagon represents the BZ of TBG.
			(c) The distribution of commensurate twist angles in TBG, where the horizontal axis is the twist angle and the vertical axis is the inverse of the number of atoms in the primitive cell (N). The larger the value, the smaller the periodicity of the system. The (q, p) values of the larger peaks are indicated.}
\end{figure}
Due to the 
$C_6$ symmetry, we only need to focus on twist angles within the range of 0 to 30 degrees. In order to more conveniently analyze the geometry of TBG, we can define a factor, 
$g=\frac{200}{N}$, to characterize the size of the periodicity of twisted bilayer graphene(TBG), where $N$ represents the number of atoms in the twisted unit cell. 
The distribution of commensurate angles can be obtained as shown in Fig. 3(c). 
It can be observed that the commensurate angles are completely discrete and exhibit a spectrum-like distribution. As the angle decreases, the length of the unit cell increases. 
By determining (q,p), a specific peak can be identified.
As an example, we show the lattice structure of TBG and the folding of its BZ at a twist angle of $21.79^\circ$ in Fig. 3.

To comprehensively demonstrate the evolution of the TBG band structure with varying twist angles, 
we selected eight structures from all possible commensurate angles. These structures, 
with the number of atoms ranging from 28 to 13,468, 
were used to calculate the band structures using the TB method, as shown in Figs. 4(a)-(h).

\begin{figure*}
	\includegraphics[width=\textwidth]{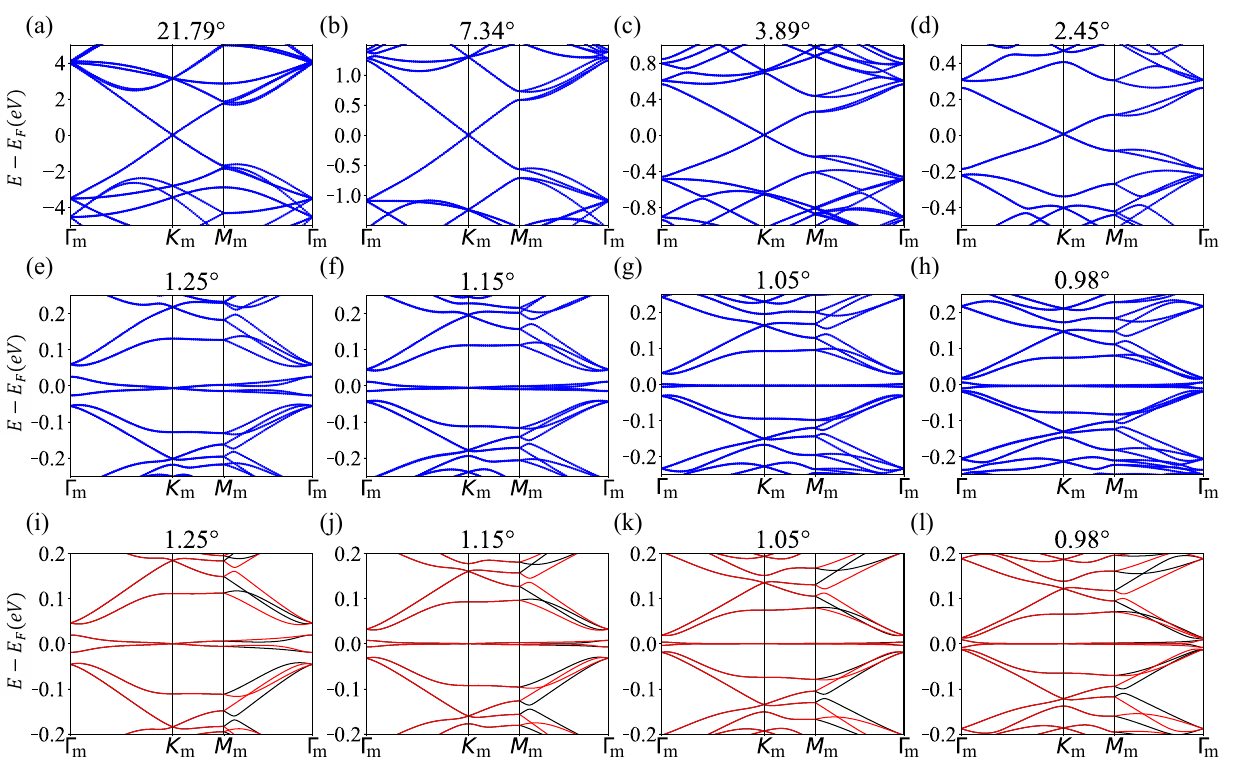}
	\caption{The band structures of TBG at different twist angles calculated using two methods are shown. 
	(a)-(h) The band structures at different twist angles were obtained using the twisted TB method. 
	(i)-(l) The band structures at small twist angles are calculated using the continuum model method, where the red and black lines represent different valleys.}
  \end{figure*}

It can be observed that as 
$\theta$ decreases, the moir\'e unit cell gradually enlarges, the periodicity extends, 
and band folding occurs, resulting in denser bands.  
At the M point, the influence of interlayer coupling on the band structure can be directly observed,
a splitting occurs in both the conduction and valence bands at the 
M point. 
Moreover, massless Dirac cones persist at the K point.  As 
$\theta$ decreases, the overall band structure is "pushed" towards the Fermi surface by the moir\'e potential, 
resulting in a reduction in Fermi velocity and consequently flatter bands.
Another noteworthy point is that the degeneracy at the $\Gamma$ point between the four low-energy bands near the Fermi level and the higher energy bands is lifted at around 2.45 degrees, 
resulting in four isolated bands, shown in Fig. 4(d). In this phenomenon, lattice relaxation plays a crucial role.
As the twist angle continues to decrease, the bandwidth of the low-energy bands keeps shrinking. 
Around 1.05 degrees, the bandwidth reaches its minimum, resulting in a completely isolated flat band with a gap of more than 10 meV from other bands on both the electron and hole sides, shown in Fig. 4(g). 
The extremely narrow bandwidth indicates that the electronic properties of TBG at this twist angle are significantly influenced by electron-electron interactions.

As shown in Fig. 1, our method allows us to study the electronic properties of twisted systems from different perspectives. 
Therefore, we proceed to construct the continuum model for TBG.
We focus on the properties near the graphene BZ 
$\pm \bm{K}$ points. Since there are gapless Dirac points, we can neglect $U_l(\bm{r})$. The continuum model for TBG can then be expressed as\cite{bistritzer_moire_2011}:
\begin{equation}
	H(\bm{r})
	=
	\left(
		\begin{array}{cc}
			h_+(-i\bm{\nabla}) & T(\bm{r}) \\
			T^{\dagger}(\bm{r}) & h_-(-i\bm{\nabla})
		\end{array}
	\right)
\end{equation}
Because there are sublattice degrees of freedom here, $H(\bm{r})$ is actually a $4\times 4$ matrix, where
\begin{equation}
	h_l(-i\bm{\nabla}) = -\hbar v [R(l \theta/2)(-i\bm{\nabla}- \bm{\kappa}_l)] \cdot (\bm{\sigma}_x,\bm{\sigma}_y)
\end{equation}
\begin{equation}
	\begin{aligned}
	T(\bm{r}) 
&=
\left(
	\begin{array}{cc}
		\delta  &  \delta^{'} \\
		\delta^{'} & \delta
	\end{array}
\right)\\
&+
\left(
	\begin{array}{cc}
		\delta  &  \delta^{'}\omega^{-1} \\
		\delta^{'}\omega  &  \delta
	\end{array}
\right) e^{i\bm{G}_1^M \cdot \bm{r}}\\
&+
\left(
	\begin{array}{cc}
		\delta  &  \delta^{'}\omega \\
		\delta^{'}\omega^{-1}  &  \delta
	\end{array}
\right) e^{i(\bm{G}_1^M+\bm{G}_2^M) \cdot \bm{r}}
	\end{aligned}
\end{equation}
Here, $\omega=e^{2\pi i/3}$. Through first-principles calculations, we determine $\hbar \nu = 5.253\ {\rm eV \cdot \AA} , \delta=0.0813\ {\rm eV}$ and $\delta^{'}=0.0951 \ {\rm eV}$.
This is close to the generally accepted parameters\cite{koshino_maximally_2018,bistritzer_moire_2011}.
Solving the Hamiltonian using the plane wave expansion method yields the band structure as shown in Figs. 4(i)-(l).
It can be seen that within the framework of our method, the band structures obtained from the continuum model and the TB method for TBG show a high degree of consistency.

Combining the results from the TB method and continuum model method for TBG, 
we can summarize the general trend of the band structure of TBG as the twist angle $\theta$ decreases. 
The Dirac points in TBG persist as $\theta$ decreases. However, due to the increasing periodicity of the moir\'e potential, 
the low energy bands are gradually pushed towards the Fermi level until they become flat. 
Additionally, there is a process where the gaps between the central 
bands and the excited bands on the electron and hole sides initially open and then gradually close. 
These two processes together lead to the emergence of isolated flat bands near the magic angle.

\subsection{B. Transition Metal Dichalcogenides}
In recent years, transition metal dichalcogenides (TMD) have provided us with various possibilities 
for stacked bilayer systems and have stimulated a substantial amount of experimental and theoretical research to help us 
understand the moir\'e properties of semiconductor materials. Here, we begin with the most common TMD material, ${\rm MoS_2}$, 
to investigate its band structure properties at large twist angles.
Monolayer $\rm MoS_2$ in 2H phase forms a hexagonal lattice with $C_3$ rotational symmetry and consists of three sublayers, as shown in Fig .5.
By stacking two layers of ${\rm MoS_2}$ in an R-type AA stacking (i.e., aligning the Mo and S atoms of the upper and lower layers), 
performing DFT simulations, 
we obtained layer spacings of $3.52\ {\rm \AA}$ and $2.86\ {\rm \AA}$ at the high-symmetry displacement points $\tau=(0,0)$ and $\tau=(1/3,1/3)$, respectively.
By rotating at a certain commensurate angle centered at the Mo point, 
we can obtain the structure as shown in Fig. 5(c).
\begin{figure}[t]
	\centering
	\includegraphics[scale=0.425]{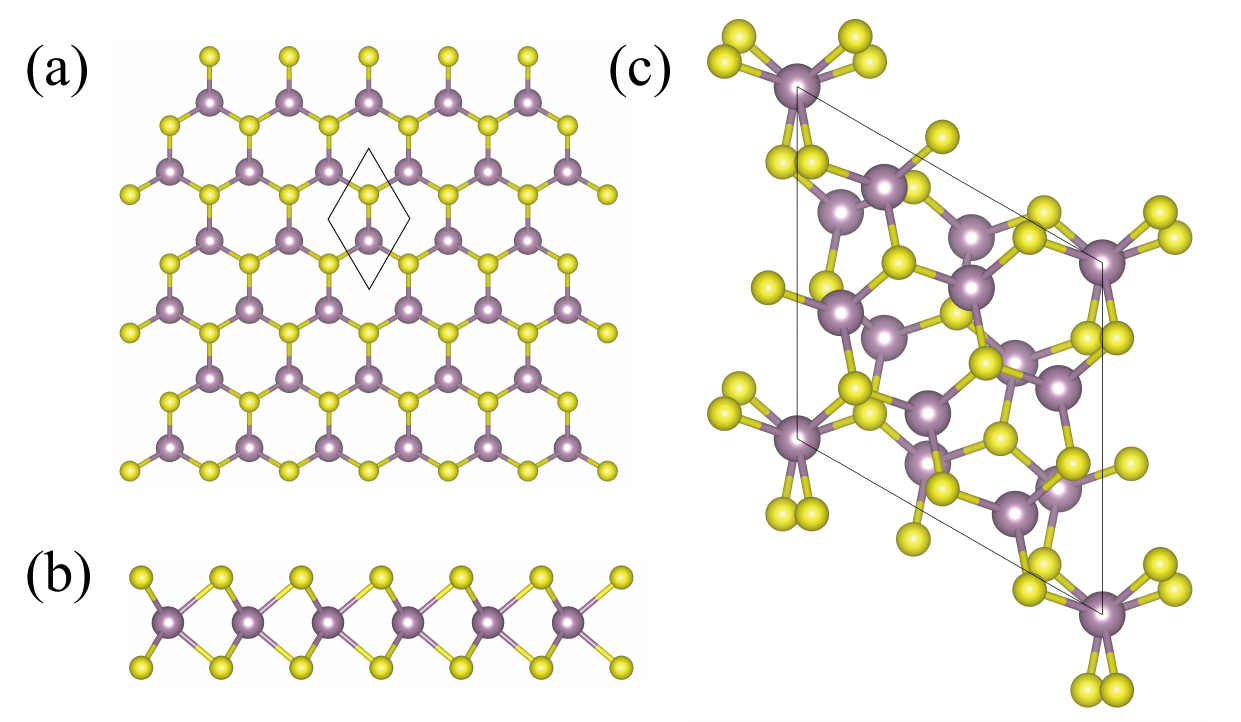}
	\caption{(a) Top view of the monolayer TMD
	structure. (b) Side view of the monolayer TMD. (c) Structure of twisted bilayer TMD at a twist angle of $21.79^\circ$. The black lines indicate the unit cell of the system.}
\end{figure}
Based on the obtained twisted structure and the TB model of monolayer ${\rm MoS_2}$, 
with the form of interlayer coupling as given in Eq. (13), 
we can calculate the band structures of several larger commensurate angles, as shown in Fig. 6.
We can observe that for the bands closest to the Fermi level, 
the energies near certain points in the BZ are almost independent of the twist angle $\theta$. 
For instance, the energy at the $\bm{K}_m$ point in the conduction band and the energy at the $\bm{\Gamma}_m$ point in the valence band are essentially unaffected by the twist angle $\theta$. 
In other words, the overall bandgap of the system remains constant for $\theta$ at large twist angles. 
However, the absolute values of their effective mass at these points increase as $\theta$ decreases,
and all bands tend to approach Fermi energy as $\theta$ decreases.  
Due to the effects of band folding and the moir\'e periodic potential, 
the valence and conduction bands become increasingly flat and separate from other bands, as shown in Fig. 6(d). 
Such isolated flat bands also indicate a significant influence on electron-electron interactions.

Next, we turn our attention to another remarkable TMD material, ${\rm MoTe_2}$, 
which exhibits unique electronic properties in the case of small twist angles, 
featuring isolated flatbands with a nonzero Chern number. 
We employ the continuum model method to investigate the electronic properties of ${\rm MoTe_2}$. 
We focus on the properties near the $\bm{K}$ point in the BZ of monolayer ${\rm MoTe_2}$.
Considering its special spin-valley locking property, 

we are specifically interested in the electronic properties of spin-up electrons within the narrow energy 
range near the Fermi level near $+\bm{K}$. For the highest valence band, following the method introduced earlier, 
firstly, we obtain the dependence of its Hamiltonian on the interlayer displacement. The form of this Hamiltonian is given by Eq. (15).

\begin{figure}[t]
	\centering
	\includegraphics[scale=0.425]{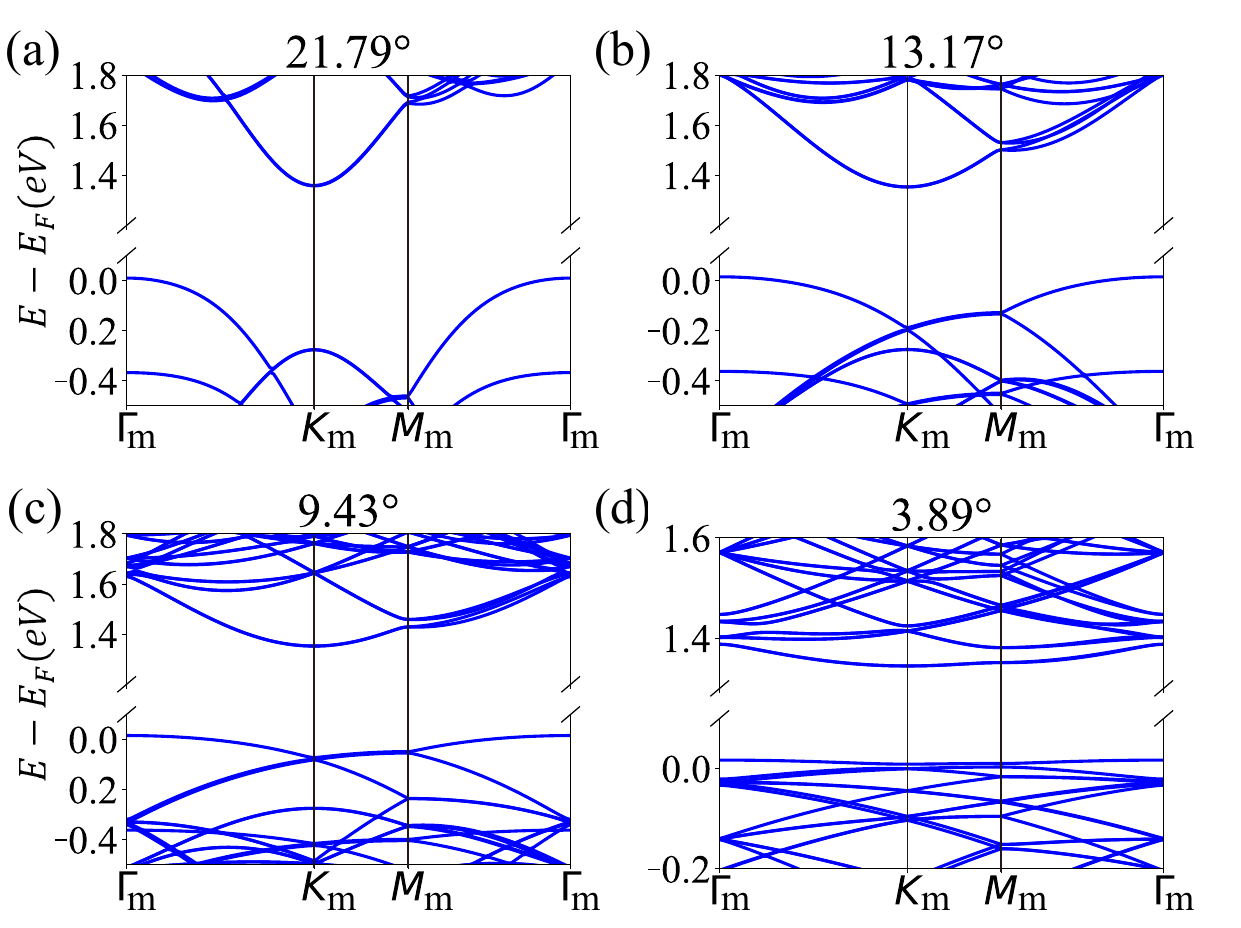}
	\caption{(a) - (d) The energy band structures of ${\rm MoS_2}$ obtained from the TB model for relatively larger commensurate angles of $21.79{^\circ}$, $13.17^{\circ}$, $9.43^{\circ}$, and $3.89^{\circ}$.}
\end{figure}

Given that the system has a hexagonal lattice structure and we are interested in the $K$ point, 
we find that the interlayer coupling is a periodic function of the interlayer displacement $\bm{\tau}$ and follows the same form as Eq. (16).
Due to the presence of $C_{3z}$ and $C_{2x}$ symmetries in the system, the intralayer influence must be invariant under threefold rotational operation and satisfy the condition $U_+(\bm{\tau}) = U_-(\bm{-\tau})$. 
Therefore, we can easily obtain the form of the intralayer influence as the form of Eq. (17).
In original bilayer ${\rm MoTe_2}$, the interlayer coupling can be characterized by band splitting, 
while the intralayer influence is manifested as a periodic influence on the overall energy band. 
Thus, we can calculate specific parameters from several high-symmetry interlayer displacements in the original bilayer structure, 
through DFT calculations. 
The values are $m^{*}=0.62m_e,\delta =-7.7\ {\rm meV}, \xi = 8.5\ {\rm meV}, \phi=-89^{\circ}$, where $m_e$ is the electron bare mass.
These parameters are very close to those in Ref.\cite{wu_topological_2019}.
With these parameters, we have determined the variation of the coupling energy with interlayer displacement. 
We still assume upper and lower layers are rotated by $\pm\theta/2$, the twisted Hamiltonian is obtained as shown in Eq. (20).
By applying the plane wave expansion method for 
twist angles of $\theta=2^{\circ}$ and $\theta=1.2^{\circ}$, we can obtain the band structures as shown in Fig 7.
\begin{figure}[t]
	\centering
	\includegraphics[scale=0.425]{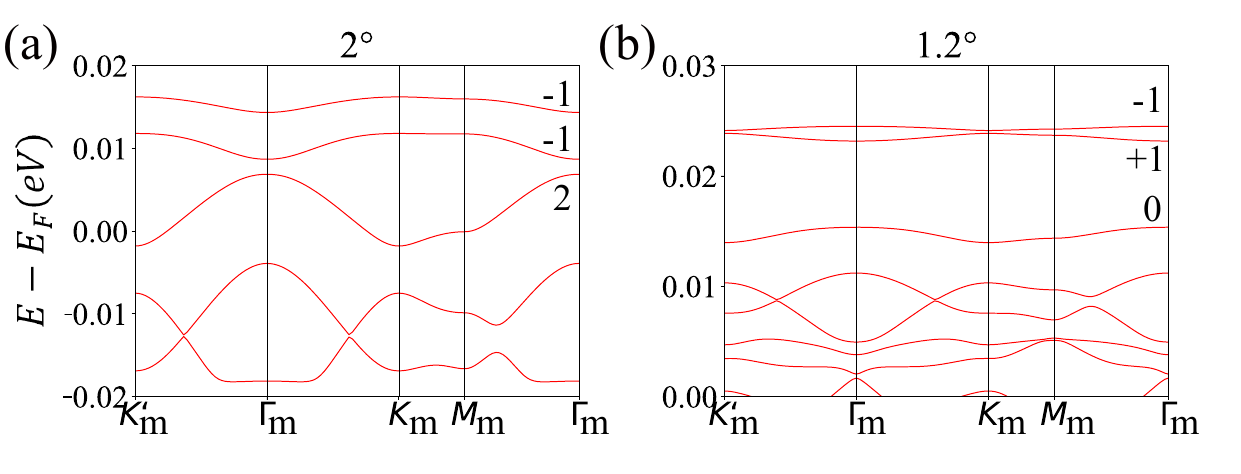}
	\caption{(a) The band structure of twisted ${\rm MoTe_2}$ with $\theta=1.2^\circ$ obtained using the kp effective model method.
    (b) The band structure of twisted ${\rm MoTe_2}$ with $\theta=2^\circ$ obtained using the kp effective model method.
    In both cases, completely isolated flat bands can be observed. The numbers in the figure indicate the Chern numbers of the bands.}
\end{figure}
We can see that in the case of small twist angles, the bandwidth of certain bands becomes very small. 
The Chern numbers of the top three bands are (-1, -1, 2) at $2^\circ$ and (-1, +1, 0) at $1.2^\circ$. 
Therefore, we can conclude that the system exhibits topologically nontrivial flat bands at small twist angles, 
and a topological phase transition occurs as the twist angle decreases.

Next, we focus on the lattice relaxation of the system. 
By combining machine learning with molecular dynamics using the software DeepMD\cite{wang_deepmd-kit_2018}, 
we can obtain precise molecular dynamics potentials, which allow us to perform accurate molecular dynamics simulations\cite{thompson_lammps_2022}.
As shown in Fig. 8, it can be observed that due to lattice relaxation, 
the MM stacking region of the system decreases, while the MX and XM regions increase. Moreover, 
distinct boundaries, referred to as domain walls, form between MX and XM regions. 
This observation is highly consistent with experimental observations\cite{shabani_deep_2021,weston_atomic_2020}.

\subsection{C. Black phosphorene}

Ever since black phosphorene was isolated from its 3D counterparts, this new generation of 2D material has attracted significant attention due to its high carrier mobility, 
layer-dependent bandgap modulation, and other excellent semiconductor properties. 
The lattice structure of black phosphorene, as shown in Fig. 9(a), consists of two sublayers arranged in a rectangular lattice. 
Black phosphorene exhibits pronounced anisotropy, with different effective masses of the conduction band minimum (CBM) and valence band maximum (VBM) along the X and Y directions.
It is worth noting that there has been limited research on twisted bilayer black phosphorene(TBP), and due to its rectangular lattice structure, thus twisted structures of black phosphorene are fundamentally different from hexagonal structures like graphene. 
There has been no systematic discussion on this matter. Furthermore, previous approaches to the structure of TBP often encountered issues. 
For instance, in dealing with large-angle TBP, researchers commonly employed methods that apply slight strains to one of the layers to approximate the commensurate structure. 
However, this approach may influence the electronic structure of the system.
The method described in this paper allows for the precise determination of the commensurate angles of TBP and the periodicity of its structure at each twist angle.

\begin{figure}[t]
	\centering
	\includegraphics[scale=0.38]{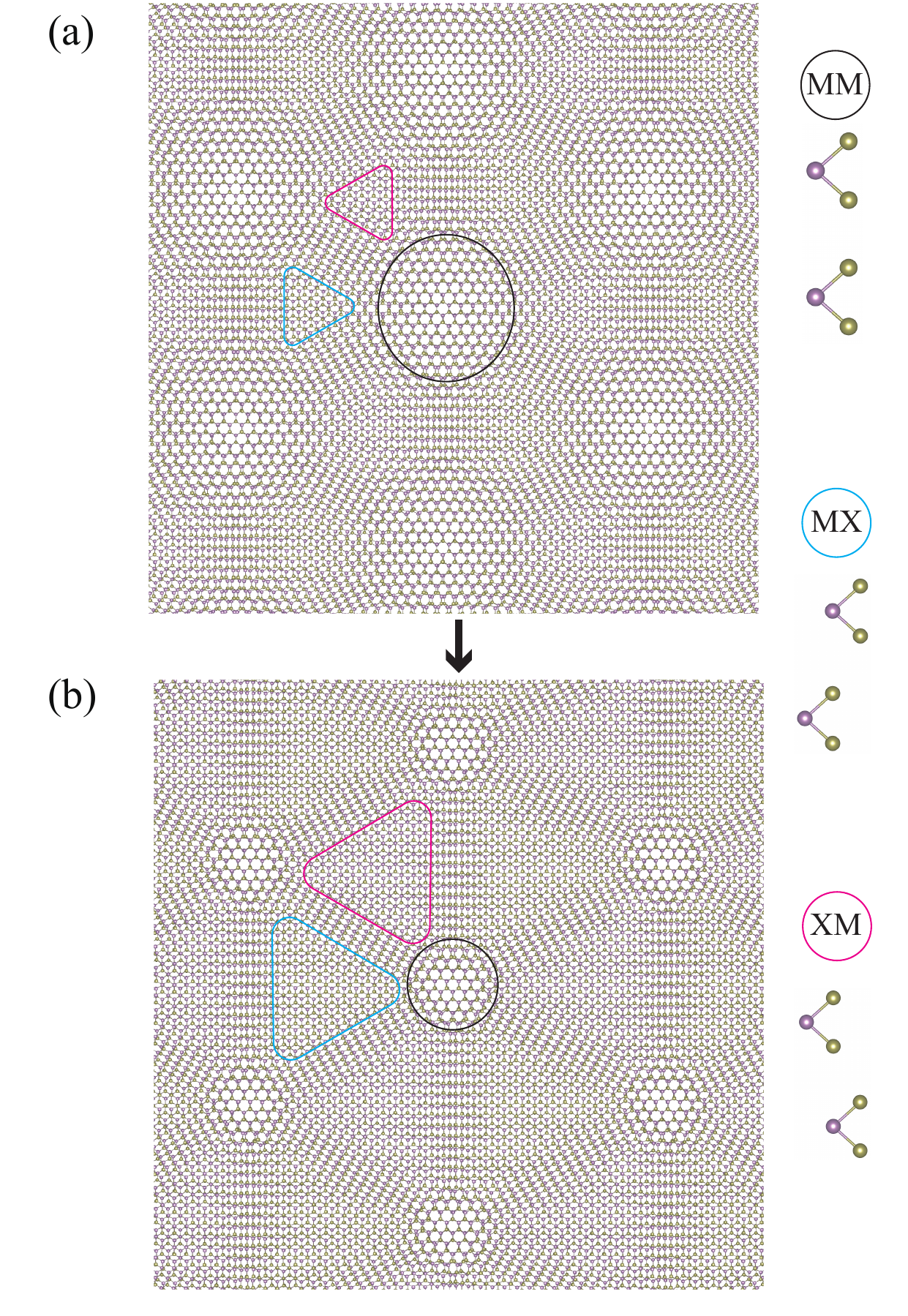}
	\caption{At twist angle of 2.88 degrees, (a) the unrelaxed structure of twisted bilayer TMD, and (b) the structure of twisted bilayer TMD after precise relaxation by molecular dynamics.
	The black, blue, and red lines represent the MM, MX, and XM stacking areas respectively.}
\end{figure}

First, focus on the structure of TBP,
The lattice vectors of black phosphorene are $\bm{a}_1 = a_0(1, 0),\bm{a}_2 = a_0(0, c), c=1.3233$,
substituting into Eq. (3) we can get
\begin{equation}
	(L_b^T)^{-1}R(\theta)L_t^T 
= 
\left(
	\begin{array}{cc}
		\cos \theta & -c\sin \theta \\
		\frac{\sin \theta}{c} & \cos \theta
	\end{array}
\right).
\end{equation}
As before, each term is required to be a rational number,
\begin{equation}
	\sin \theta = \frac{1}{c}\frac{l_1}{l_3},
    \cos \theta = \frac{l_2}{l_3},
\end{equation}
and we have the Diophantine equation
\begin{equation}
	\frac{1}{c^2}l_1^2+l_2^2=l_3^2,
\end{equation}
where
\begin{equation}
	l_1=2qp
,l_2=\frac{1}{c^2}q^2-p^2
,l_3=\frac{1}{c^2}q^2+p^2,
\end{equation}
the commensurate angle is described as
\begin{equation}
	\theta = \arccos (\frac{q^2-c^2p^2}{q^2+c^2p^2})
\end{equation}
Therefore, we can also describe the structure of the twisted system through a set of positive integers.
We get the commensurate angle distribution of TBP, as shown in Fig. 9(d),
\begin{figure}[t]
	\centering
	\includegraphics[scale=0.43]{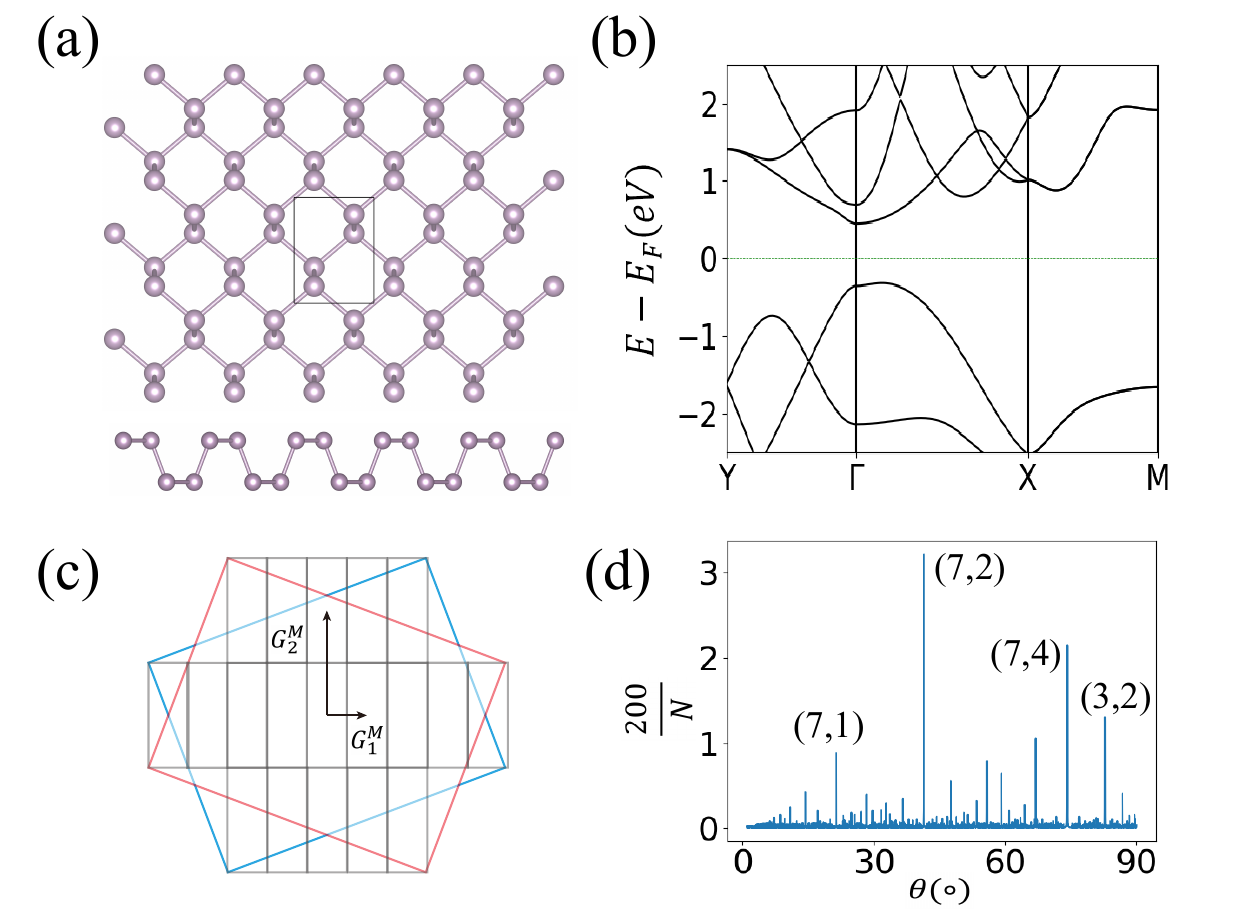}
	\caption{(a) The structure of black phosphorene. (b) The band structure of monolayer black phosphorene.
	(c) BZ folding of twisted black phosphorene, taking $41.40^{\circ}$ TBP as an example. The red (blue) rectangle represents the BZ of the original bilayer structure rotated by an angle of $\frac{-\theta}{2}$($\frac{\theta}{2}$), 
	and the black rectangle represents the BZ of twisted black phosphorene.
	(d) The distribution of commensurate twist angles in twisted black phosphorene, where the horizontal axis is the twist angle and the vertical axis is the inverse of the number of atoms in the primitive cell (N). 
	The larger the value, the smaller the periodicity of the system. The (q, p) values of the larger peaks are indicated.}
\end{figure}
similarly exhibiting a spectrum-like distribution, the commensurate angles distribution of black phosphorus is more disordered compared to hexagonal structures. However, 
they can still be categorized into different peaks based on different 
$q$ and 
$p$ values.
\begin{figure}[htbp]
	\centering
	\includegraphics[scale=0.425]{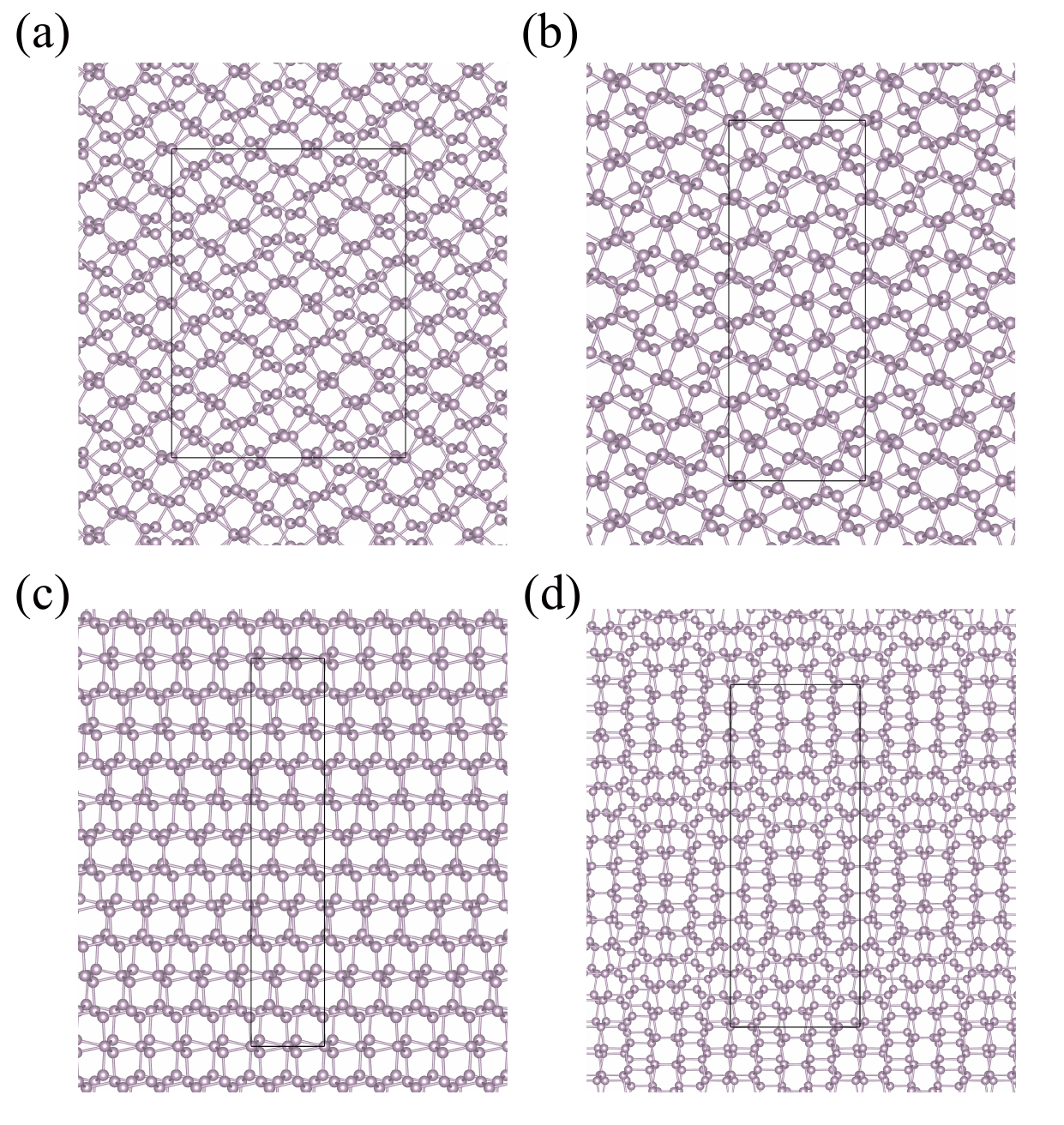}
	\caption{Four commensurate twisted black phosphorene structures. (a)-(b) are the structures at twist angles of $21.40^{\circ},41.40^{\circ}$,
	$74.16^{\circ}$ and $82.80^{\circ}$ respectively. The black line area is the moir\'e unit cell.}
\end{figure}

We have selected four commensurate structures, as shown in Fig. 10.
For a rectangular lattice 2D system like black phosphorus, 
its twisted structures will also be either rectangular lattices or centered rectangular lattices. 
At different twist angles, the system exhibits varying degrees of anisotropy. 
For the four structures shown in Fig. 10, the aspect ratios of their lattice vectors are 1.32, 2.65, 5.29, and 2.65, 
respectively. Therefore, we consider twisting as an effective means of modulating anisotropy.
Additionally, in highly anisotropic twisted structures, certain degrees of one-dimensional (1D) pattern can be observed, 
as seen in Figs. 10(b) and 10(c). This results in distinctive electronic structures, highlighting their potential for simulating 
1D systems.

Next, we investigate the electronic structure of TBP to substantiate these points. 
We obtained the TB Hamiltonian and band structure for the four structures mentioned above, as shown in Fig. 11.
\begin{figure}[htbp]
	\centering
	\includegraphics[scale=0.425]{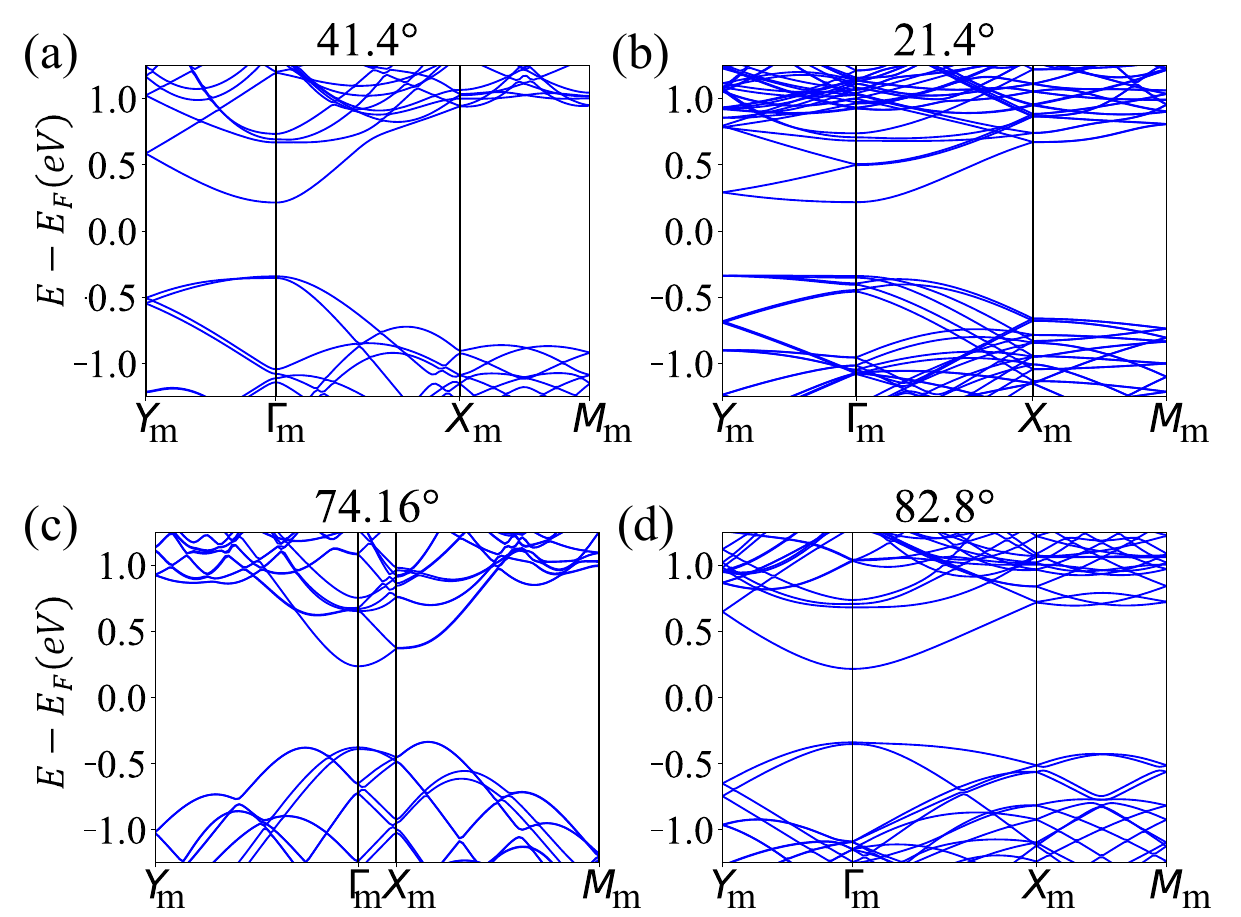}
	\caption{(a)-(d), band structures obtained by the TB method at twist angles of $41.40^{\circ},21.40^{\circ},$,
	$74.16^{\circ},82.80^{\circ}$.}
\end{figure}

It can be observed that the band structure of TBP also exhibits strong anisotropy. 
For example, in Fig. 11(b), the valence and conduction bands have completely different bandwidths along the $\Gamma_m-Y_m$ and $\Gamma_m-X_m$ paths. 
In the $\Gamma_m-Y_m$ direction, the low-energy bands are almost non-dispersive, indicating that the wavefunctions at the $\Gamma$ point have no overlap in the y direction but do touch in the x direction, 
showing distinct quasi-one-dimensional characteristics.

Additionally, it is worth mentioning that monolayer black phosphorus is an approximately direct bandgap semiconductor, 
with the conduction band minimum located at the BZ center ($\Gamma$ point) and the valence band maximum slightly 
offset from the $\Gamma$ point (by about $2\%$). This small offset is usually negligible, but in moir\'e black phosphorus, 
due to band folding, this offset is amplified. As shown in Fig. 11(c), the VBM is already far from the $\Gamma_m$ point. 
Therefore, through twisting, black phosphorus transforms from a direct bandgap semiconductor to an indirect bandgap semiconductor.

Furthermore, we focus on the effective mass of TBP, as shown in Fig. 12. 
The effective mass of the valence and conduction bands is highly dependent on the twist angle, 
both in magnitude and direction. Notably, when the twist angle is $82.80^{\circ}$, 
the angle between the directions of the maximum effective mass of VBM and CBM is $90^{\circ}$,
meaning the directions of the greater dispersion in the valence and conduction bands are perpendicular to each other. 
This uniquely tunable anisotropy may have potential research value in physical property.
\begin{figure}[t]
	\centering
	\includegraphics[scale=0.45]{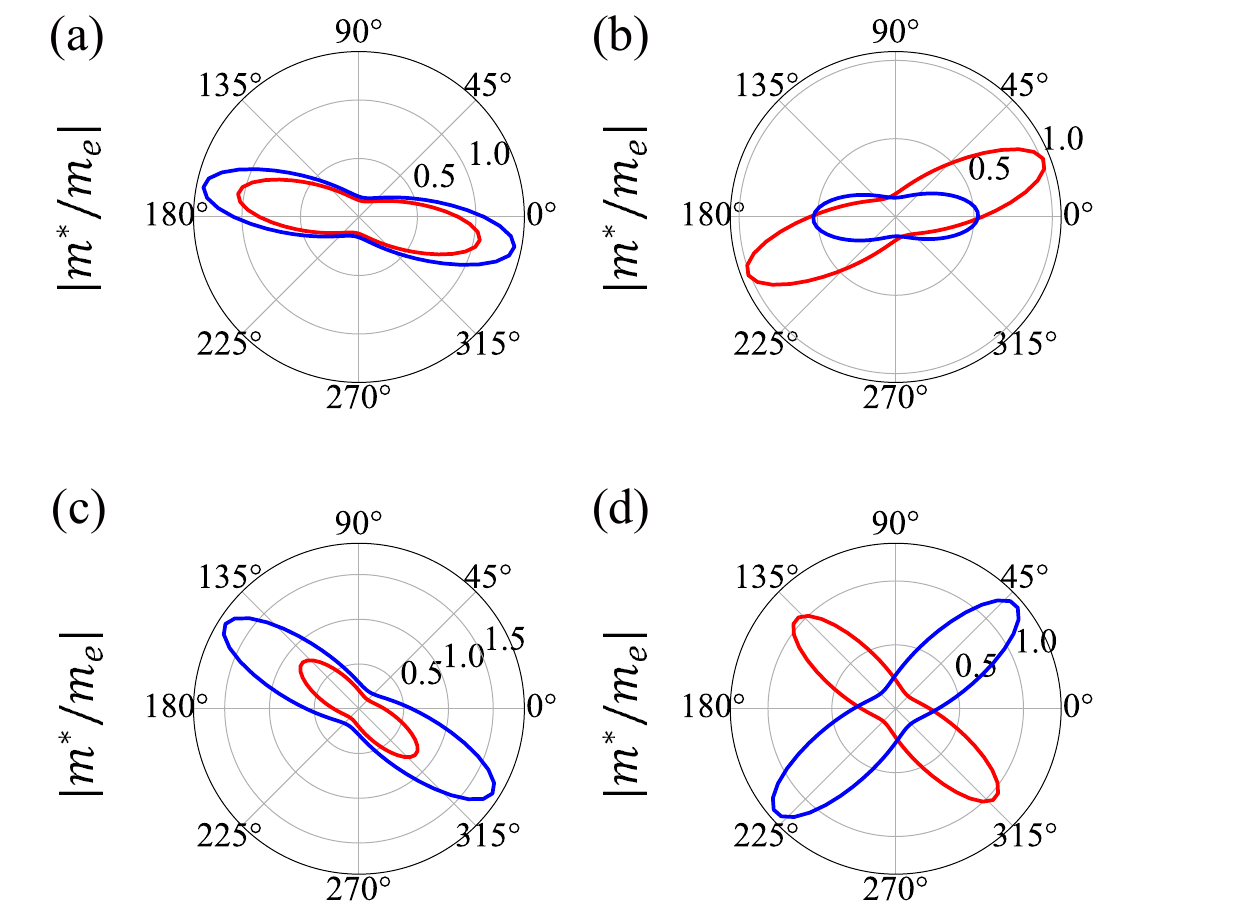}
	\caption{The absolute value of the effective mass of VBM (red line) and CBM (blue line) of TBP varies with the direction in BZ under different twist angles. 
	(a)-(b) represent the cases where the twist angles are $21.40^{\circ}, 41.40^{\circ}$, $74.16^{\circ}, 82.80^{\circ}$ respectively.}
\end{figure}

For small twist angle black phosphorene, we can also study its electronic properties through the kp-continuum model method.
Since we focus on the properties of the valence band near the $\Gamma$ point of the system and consider the rectangular lattice properties of the system,
we can get its continuum model Hamiltonian as
\begin{equation}
	H(\bm{r})
=
\left(
	\begin{array}{cc}
		h(-i\bm{\nabla})  &  T(\bm{r}) \\
		T^{\dagger}(\bm{r})  &  h(-i\bm{\nabla})
	\end{array}
\right),
\end{equation}
the intralayer part of the Hamiltonian is
\begin{equation}
	h(-i\bm{\nabla}) = \frac{\hbar^{2}(-i\bm{\nabla} \cdot \bm{e}_{x})^2}{2m_{x}^{*}} + \frac{\hbar^{2}(-i\bm{\nabla} \cdot \bm{e}_{y})^2}{2m_{y}^{*}},
\end{equation}
The effect of interlayer displacement on the intralayer Hamiltonian is very small and can be ignored. The interlayer coupling term can be written as
\begin{equation}
	T(\bm{r}) = \delta_0 + \delta_1e^{i\bm{G}_1^M \cdot \bm{r}} + \delta_2e^{i\bm{G}_2 ^M\cdot \bm{r}} 
\end{equation}
The parameters are still calculated by DFT, $m_x=-0.17m_e, m_y=-1.12m_e$, $(\delta_0, \delta_1, \delta_2)=(10,140,12) meV$,
The above Hamiltonian is solved by the plane wave expansion method to obtain the band structure shown in Fig. 13.
\begin{figure}[t]
	\centering
	\includegraphics[scale=0.425]{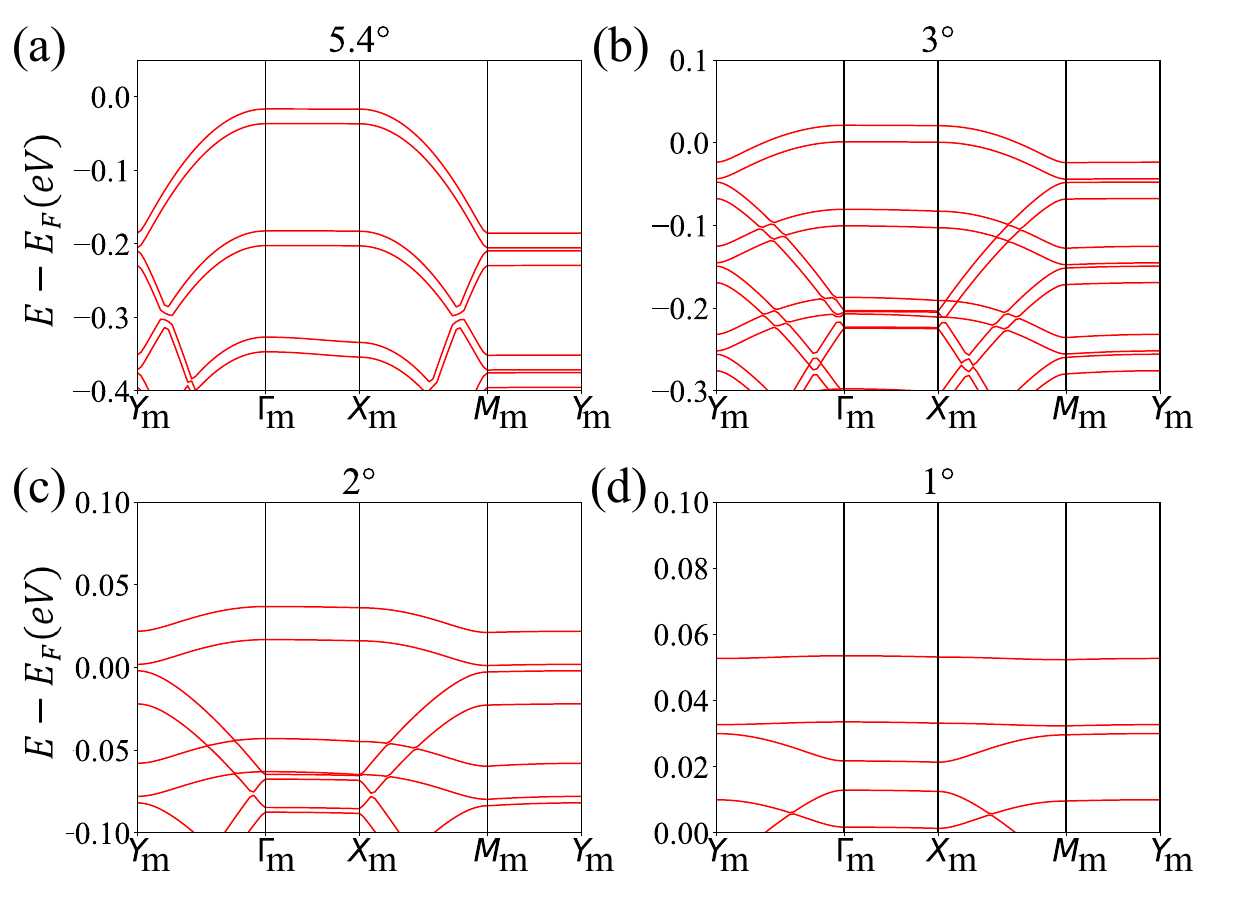}
	\caption{The continuous model band structure of twisted black phosphorene (a)-(d), with twist angles of $5.4^{\circ},3^{\circ}$,
	$2^{\circ},1^{\circ}$ respectively.}
\end{figure}
It can be observed that the band structure obtained from the continuum model shows that for larger twist angles, 
the valence band is nearly non-dispersive along the $\Gamma_m-X_m$ path but highly dispersive along the $\Gamma_m-Y_m$ direction(as shown in Fig. 13(a)). 
This indicates that the system maintains a high degree of anisotropy. As the twist angle decreases, 
the valence band gradually separates from the other bands and its bandwidth decreases. 
When the twist angle reaches $1^{\circ}$ (as shown in Fig. 13(d)), a completely isolated flat band appears,
with a bandwidth much smaller than the gap to other bands. Due to the extremely small bandwidth and the relatively large bandgap, 
the correlation effects among electrons in this band could be very significant, 
making this system highly valuable for further research.

\begin{figure}[t]
	\centering
	\includegraphics[scale=0.425]{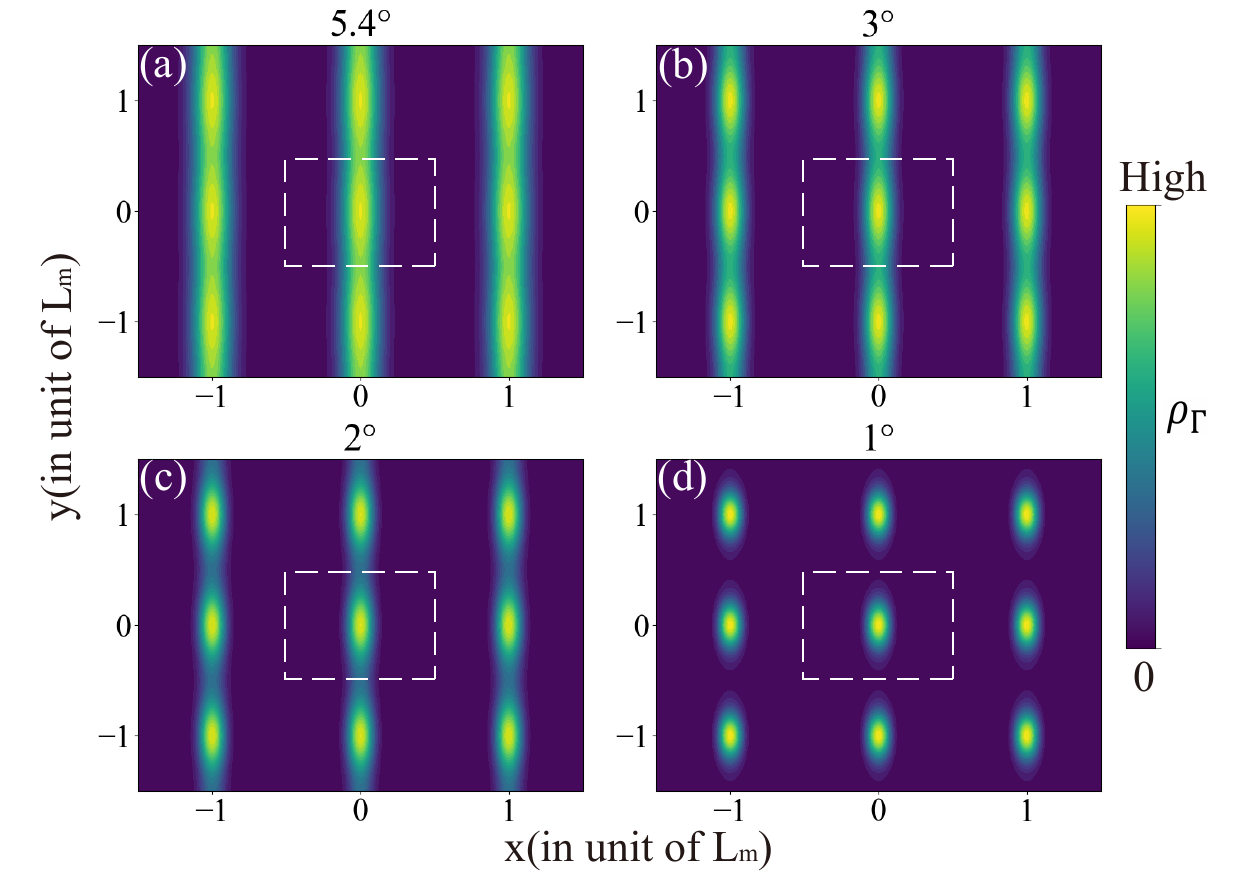}
	\caption{The modulus of the wavefunction of TBP's valence band at the $\Gamma$ point of the BZ in real space, 
	(a)-(d), with twist angles of $5.4^{\circ},3^{\circ}$,$2^{\circ},1^{\circ}$ respectively. 
	The white dashed rectangles indicate the moir\'e unit cell of the system, 
	with the horizontal and vertical axes given in units of the moir\'e lattice vectors $L_m$.}
\end{figure}

The unique anisotropic band structure of TBP directs our attention to the real-space distribution of its valence band wavefunction modulus, 
as shown in Fig. 14. The real-space distribution of the wavefunction at the $\Gamma$ point of TBP's BZ is highly influenced by the twist angle. 
At larger twist angles, corresponding to the band dispersion characteristics of the system, 
the wavefunction distribution shows no overlap in the x direction but has significant overlap in the y direction, 
forming completely one-dimensional stripes. This demonstrates TBP's great potential to simulate one-dimensional systems.
As the twist angle decreases, the overlap of the wave function in the y direction decreases, and it begins to become more and more localized, 
the anisotropy weakens, eventually leading to the formation of a flat band.

In this section, we followed the workflow shown in Fig. 1 to study the properties of TBP, 
from its structure to the TB model and continuum model. 
We found that the anisotropy of TBG is highly tunable with respect to the twist angle. 
During the variation of the twist angle, interesting physical characteristics may emerge, such as anisotropic dispersive bands, 
uniquely shaped effective mass of the valence and conduction bands, isolated flat bands, 
and one-dimensional electronic characteristics,
all of which warrant further investigation.

\section{$\rm{\uppercase\expandafter{\romannumeral4}}$. discussion}
In summary, we have developed a general versatile method for the study of electronic structures in twisted systems, 
which starts from DFT calculations. 
This method allows us to generate the structure of twisted systems, 
perform TB and effective $\bm{k} \cdot \bm{p}$ continuum model calculations for electronic structures, 
and achieve accurate band structures with minimal computational cost.
It provides a solid platform for investigating twisted electronic systems. 
Currently, our method is limited to calculating the band structures of homogenous twisted materials. 
Therefore, in our future work, we plan to extend our research to include heterostructures 
and twisted transport properties. We will consider releasing our method as open-source software to the community when appropriate.

\section{$\rm{\uppercase\expandafter{\romannumeral5}}$. ACKNOWLEDGMENTS}
The work is supported by the NSF of China (Grant No. 12374055), the National Key R\&D Program of China (Grant No. 2020YFA0308800), 
and the Science Fund for Creative Research Groups of NSFC (Grant No. 12321004).
\bibliography{ref}
\end{document}